\documentclass[letterpaper,twoside,10pt]{ieeeconf}
\usepackage{generic}
\usepackage{cite}
\usepackage{amsmath,amssymb,amsfonts}
\usepackage{mathrsfs}
\usepackage{algorithmic}
\usepackage{graphicx}
\usepackage{tikz}
\usepackage{pgfplots}
\pgfplotsset{compat=1.18}
\usepackage{empheq}
\usepackage{comment}
\newtheorem{theorem}{Theorem}
\newtheorem{proposition}{Proposition}
\newtheorem{corollary}{Corollary}
\newtheorem{property}{Property}
\newtheorem{remark}{Remark}
\newtheorem{example}{Example}
\newtheorem{definition}{Definition}
\newtheorem{ass}{Assumption}
\newtheorem{lem}{Lemma}

\newenvironment{Theorem}{\begin{theorem}}{\hfill $\square$ \end{theorem}}
\newenvironment{Proposition}{\begin{proposition}}{\hfill $\square$ \end{proposition}}

\newcommand{\sign}{\operatorname{sign}}
\usepackage{algorithm,algorithmic}
\usepackage{hyperref}
\hypersetup{hidelinks=true}
\usepackage{textcomp}
\begin{document}
\title{\bf Fast Sampling or High Gain? iP versus PI Control}
\author{Mohamed Camil Belhadjoudja
\thanks{The author is with Université Grenoble Alpes,
CNRS, Grenoble-INP, GIPSA-lab, F-38000, Grenoble, France (e-mail: mohamed.belhadjoudja@gipsa-lab.fr)}}

\maketitle
\pagenumbering{gobble}

\begin{abstract}
Intelligent proportional (iP) controllers, introduced in the framework of model-free control, are often presented as an alternative to proportional–integral (PI) controllers that requires no model of the plant. This paper studies this property for a first-order linear plant under sampled-data implementation with a zero-order hold. When the time derivative of the output is estimated by a backward difference, the discrete-time iP and PI controllers are known to obey the same control law under a one-to-one correspondence between their parameters. Building on this correspondence, the paper derives necessary and sufficient conditions for exponential stability of the iP closed loop, in closed form, in terms of the controller parameters $K_P$ and $\alpha$, the sampling period $h$ and the plant parameters. It then shows that, when the iP parameters are held fixed and the sampling period tends to zero, the closed-loop decay rate tends to the iP gain $K_P$, provided that $K_P > 0$ and that the ratio $\rho$ between the input gain of the plant and $\alpha$ satisfies $0 < \rho < 2$; one closed-loop pole then behaves as $1 - hK_P$ and the other tends to $1 - \rho$. With fixed PI gains, by contrast, the decay rate tends to that of the continuous-time PI loop, which depends on the plant parameters. The two parametrizations therefore suggest different tuning procedures, although they achieve the same closed loops at every fixed sampling period. High gain for the PI is, in some sense, replaced by fast sampling of the iP. The results also make explicit which plant information the iP controller requires: the sign of the input gain and a bound on its magnitude in the limit of fast sampling, and bounds on both plant parameters at a given sampling period. Numerical experiments illustrate these results.
\end{abstract}

\section{Introduction}

Proportional-integral (PI) controllers, and more generally, PID controllers, remain the standard feedback laws in industrial practice. They are almost always implemented digitally: the output is sampled with period $h$, the control is updated at the sampling instants, and held constant between them. For a first-order plant $\dot y=py+bu$, the closed loop with a sampled PI controller is a second-order linear recurrence whose exact stability region can be obtained from Jury's test \cite{landau,AW97,Ela05,Jur64}. For fixed gains, this recurrence converges as $h\to0^+$ to the continuous-time PI loop with characteristic polynomial $s^2+(bk_P-p)s+bk_i$, where $k_P$ and $k_i$ denote the proportional and integral gains, respectively. The limiting closed-loop poles therefore depend on the plant parameters $p$ and $b$, and assigning them requires some knowledge of these parameters.

Model-free control (MFC) \cite{FJ09,FJ13} was introduced to avoid the need for a plant model in controller design. The unknown input-output dynamics are represented by the ultra-local model $\dot y=F+\alpha u$, where $\alpha$ is a design constant and $F$ accounts for the remaining dynamics. The quantity $F$ is estimated online from past inputs and outputs. The intelligent proportional (iP) controller
$$
u=-\frac{\hat F+K_Pe}{\alpha},
$$
where $K_P$ is a controller gain, $\hat F$ is an estimate of $F$, and $e=y-y^*$ is the regulation error with $y^*$ constant, aims to impose the error dynamics $\dot e \approx -K_Pe$ when $\hat{F}-F \approx 0$. An accurate identification of the input gain is not considered necessary in the MFC literature \cite{FJ13,JDFM17}. 

The relation between the iP and PI controllers has been established for their discrete-time implementations. In \cite{DN10}, it was shown that, when $F$ is estimated using a backward difference of the output and the previous input, the sampled iP controller and a sampled PI controller with a Riemann-sum integral obey the same incremental law under the parameter relations
$$
k_P=\frac{1}{\alpha h},
\qquad
k_i=\frac{K_P}{\alpha h},
$$
with analogous relations between sampled PID and intelligent PD controllers. The same equivalence was recovered in \cite{La23} through a Youla-Ku\v{c}era parametrization. Fliess and Join have since advocated replacing PI and PID regulators with intelligent controllers \cite{FJ22}. What these equivalences imply for the closed-loop dynamics, and in particular what happens when $h$ tends to zero with $\alpha$ and $K_P$ held fixed, has not been studied in the literature, to the best of our knowledge.

Stability analyses of control systems under the iP controller have mainly relied on continuous-time idealizations. Delaleau \cite{De14} interprets the discrete MFC law as an explicit Euler scheme for an ordinary differential equation and proves local asymptotic tracking for that equation using singular perturbation theory. The work \cite{JDFM17} models the estimator $\hat{F}$ using derivative filters and shows, through the Routh-Hurwitz criterion, that the stabilizing parameter region can be narrow for a second-order unstable plant. In \cite{BMW23}, the estimator is modeled using the exact output derivative and a delayed input. The resulting iP loop with a linear plant is shown to be a neutral or advanced delay system, and instability results and sufficient conditions for exponential stability are derived. The work \cite{Sch25} obtains plant-dependent parameter bounds using the generalized Hermite-Biehler theorem, again for a continuous-time delay model. None of these works provides a stability analysis of a plant controlled by a discrete-time iP controller, which is the controller actually implemented in practice. The equivalence established in \cite{DN10} allows the analysis to be reduced to that of a sampled PI loop. However, the resulting stability conditions in terms of the iP parameters, as well as the behaviour of the closed-loop poles as $h\to0^+$ with $(\alpha,K_P)$ fixed, have not been characterized in the literature.

In this paper, we analyze the sampled-data iP controller for the plant $\dot y=py+bu$ under a zero-order hold, using the estimate of $F$ considered in \cite{FJ09,DN10}. We recall the correspondence established in \cite{DN10} and make three contributions.

\begin{enumerate}
\item \emph{Exact stability conditions.} We derive necessary and sufficient conditions for exponential stability of the sampled iP loop in closed form, in terms of $\rho=b/\alpha$, $K_P$, and $ph$ (Theorem \ref{thm_1}). They identify precisely what plant information is needed to certify a given choice of $(\alpha,K_P,h)$.

\item \emph{Small-sampling asymptotics.} For $0<\rho<2$ and $K_P>0$, the two closed-loop poles under iP control satisfy
$$
z_f(h)=1-\rho+O(h),\quad
z_s(h)=1-hK_P+O(h^2),
$$
where $z_f$ and $z_s$ denote the fast and slow poles, respectively. Consequently, the decay rate tends to $K_P$ as $h\to0^+$ (Theorem \ref{thm:small_h}). Conversely, outside these conditions, the loop is not exponentially stable for sufficiently small $h$.

\item \emph{Two tuning philosophies: fast sampling for the iP versus high-gain PI.} With fixed PI gains, the decay rate tends to that of the continuous-time PI loop and therefore depends on $p$ and $b$. With fixed $(\alpha,K_P)$, the decay rate tends to $K_P$ regardless of $p$, while the equivalent PI gains increase as $1/h$. Thus, although the two parametrizations yield exactly the same closed-loop characteristic polynomial at every fixed $h$, they correspond to two different tuning philosophies: increasing the PI gains, or keeping $(\alpha,K_P)$ fixed and increasing the sampling frequency. Accordingly, the plant independence of the iP tuning holds in an asymptotic
sense. The small-sampling limit requires the sign of $b$, the bound $|b|<2|\alpha|$, and $K_P>0$, whereas certifying stability at a given sampling period requires bounds on $p$ and $b$, as in conventional PI design.
\end{enumerate}

The remainder of the paper is organized as follows. Section \ref{sec:notation} introduces the notation used throughout the paper. Section \ref{background} reviews preliminary results on the stability of first-order single-input-single-output (SISO) linear time-invariant (LTI) systems under PI control and recalls the discrete-time iP control law. Section \ref{sec_inp} presents and discusses our main results. Section \ref{sec:numerics} illustrates these results through numerical experiments. Finally, concluding remarks and perspectives for future research are provided.

\section{Notation}\label{sec:notation}

\paragraph*{Numbers and sets}
$\mathbb{R}$ and $\mathbb{C}$ denote the real and complex numbers, and
$\mathbb{R}^*:=\mathbb{R}\setminus\{0\}$. The set of natural numbers is
$\mathbb{N}:=\{0,1,2,\dots\}$, and $\mathbb{N}^*:=\mathbb{N}\setminus\{0\}$.
For $z\in\mathbb{C}$, $|z|$ is its modulus and $\operatorname{Re}z$ its real part.
For $x\in\mathbb{R}$, $\operatorname{sgn}(x)\in\{-1,0,1\}$ is the sign of $x$, with the convention $\sign(0)=0$.
We denote by $\ln$ the natural logarithm, and $\coth (x):=(e^{x}+e^{-x})/(e^{x}-e^{-x})$ for
$x\in\mathbb{R}^*$.

\paragraph*{Asymptotic notation}
Let $f$ and $g$ be functions defined on an interval $(0,h_1)$, with $f$ complex-valued
and $g$ real-valued and positive. We write $f(h)=O(g(h))$ as $h\to0^+$ if there exist
constants $C>0$ and $h_2\in(0,h_1]$ such that $|f(h)|\le C\,g(h)$ for every
$h\in(0,h_2)$. An equality of the form $f(h)=a+O(h)$ means $f(h)-a=O(h)$.

\paragraph*{Polynomials, shift operator and spectral radius}
For a polynomial $P$ with real coefficients, $r(P):=\max\{|z|: z\in\mathbb{C},\
P(z)=0\}$ is its spectral radius, that is, the largest modulus of its roots. The
polynomial $P$ is called \emph{Schur} if all its roots lie in the open unit disc,
that is, if $r(P)<1$. The forward shift operator $q$ acts on a real sequence
$(x_k)_{k\in\mathbb{N}}$ by $(qx)_k:=x_{k+1}$. For a polynomial
$P(z)=\sum_{j=0}^{n}a_jz^j$, the statement $P(q)x_k=0$ means
$\sum_{j=0}^{n}a_jx_{k+j}=0$.

\section{The PI and iP controllers applied to first-order SISO LTI systems}\label{background}

The material of Sections \ref{sec_1}-\ref{sec_3} is classical and can be found in the textbooks \cite{landau,AW97,Ela05,Jur64}.

\subsection{First-order SISO LTI systems}\label{sec_1}
In this article, we consider linear systems with control input $u(t)\in \mathbb{R}$ and output $y(t)\in \mathbb{R}$, modeled by first-order dynamics 
\begin{align}
\dot{y} = py + bu, \quad y(0)=y_0, \label{first}
\end{align}
where $p\in \mathbb{R}$, $b\in \mathbb{R}^*$, and $y_0\in \mathbb{R}$ is the initial condition. 

The controller is sampled with a zero-order hold. Specifically, given a sampling step $h \in (0,+\infty)$, we consider the sequence of time instants $(t_k)_{k\in \mathbb{N}} \subset [0,+\infty)$ defined by 
\begin{align}
t_k := hk \qquad \forall k\in \mathbb{N}. 
\end{align}
Then, the controller is held constant on each time interval $[t_k,t_{k+1})$, i.e., 
\begin{align}
u(t) = u_k := u(t_k) \qquad \forall t\in [t_k,t_{k+1}).
\end{align}
As a consequence, \eqref{first} can be transformed into a discrete-time linear control system. Indeed, integrating \eqref{first} and using the fact that $u$ is constant on $[t_k,t_{k+1})$, we obtain 
\begin{align}
y_{k+1} = \bar{p}y_k + \bar{b}u_k \qquad \forall k\in \mathbb{N}, \label{first_d1}
\end{align}
where $y_k := y(t_k)$, and 
\begin{align}
&\bar{p} := e^{ph}, \qquad \bar{b} := b\int_0^h e^{p(h-\theta)}d\theta. \label{first_d2}
\end{align}

This system can be written in the compact form 
\begin{align}
A_h(q)y_k = B_h(q)u_k \qquad \forall k\in \mathbb{N}, \label{general_system}
\end{align}
where 
\begin{align}
A_h(q) := q-\bar{p}, \quad B_h(q) := \bar{b}.
\end{align}

\subsection{Characteristic polynomial and exponential stability}

Let $y^*$ be a constant reference, and define the regulation error 
\begin{align}
e_k := y_k -y^* \qquad \forall k\in \mathbb{N}. 
\end{align}
The objective is to design the control input to achieve exponential stability of the origin $\{e=0\}$. The controller, whether it is a PI or an iP, will take, as we will see later, the form 
\begin{align}
(q-1)u_k = - K(q) e_k \qquad \forall k\in \mathbb{N},\label{controller}
\end{align}
where $K(q)$ is some polynomial in $q$ of degree at most $1$. As a result, by applying $(q-1)$ at both sides of \eqref{general_system}, while using the fact that $(q-1)y^*=0$ since $y^*$ is constant, we obtain 
\begin{align*}
(q-1)A_h(q)y_k &= (q-1)B_h(q)u_k \\
A_h(q)(q-1)y_k &= B_h(q)(q-1)u_k \\
A_h(q)(q-1)y_k - A_h(q)(q-1)y^* &= -B_h(q)K(q)e_k \\
A_h(q)(q-1)e_k &= -B_h(q)K(q)e_k. 
\end{align*}
Hence, 
\begin{align}
\chi (q) e_k = 0 \qquad \forall k\in \mathbb{N}, 
\end{align}
where 
\begin{align}
\chi(q) := A_h(q)(q-1)+B_h(q)K(q)
\end{align}
is the \textit{characteristic polynomial} of the closed-loop system. The roots of $\chi$, i.e., the elements of the set 
$$\{z\in \mathbb{C} : \chi(z)=0\},$$
are referred to as the \textit{poles} of the closed-loop system. Furthermore,
\begin{align}
r_\chi := \max\left\{|z| \, : \, z\in\mathbb{C}, \ \chi(z)=0\right\}\label{spc_radius}
\end{align}
denotes the \textit{spectral radius} of $\chi$.

Since $K(q)$ and $B_h(q)$ are of degree at most $1$, and $A_h(q)$ is of degree $1$, then $\chi(q)$ is of degree $2$. Specifically, we have 
\begin{align}
\chi(q) = q^{2} + c_1q + c_0, 
\end{align}
for some constants $(c_0,c_1)\in \mathbb{R}^{2}$. As a result, the closed-loop system $\chi(q)e_k=0$ reads
\begin{align}
e_{k+2} = -c_1e_{k+1}-c_0e_k \qquad \forall k\in \mathbb{N}, \label{recurrence}
\end{align}
so that the sequence $(e_k)_{k\in\mathbb{N}}$ is uniquely determined by its initial condition $(e_0,e_1)\in\mathbb{R}^{2}$.

\begin{definition}\label{def_stab}
The equilibrium $\{e=0\}$ for \eqref{recurrence} is exponentially stable if there exist constants $C>0$ and $r\in (0,1)$ such that, for every $(e_0,e_1)\in \mathbb{R}^2$, 
\begin{align}
|e_k| \leq Cr^k \max_{j\in \{0,1\}}|e_j| \qquad \forall k\in \mathbb{N}. \label{decay}
\end{align}
\hfill $\bullet$
\end{definition}

The exponential stability of $\{e=0\}$ is determined by the spectral radius $r_{\chi}$, as recalled in the following result. 

\begin{Proposition}\label{prop_stab}
Let $r_\chi$ be the spectral radius in \eqref{spc_radius}. 
\begin{itemize}
\item If $r_\chi<1$, then, for each $r\in(r_\chi,1)$, there exists $C>0$ such that \eqref{decay} holds for each initial condition $(e_0,e_1)\in\mathbb{R}^{2}$. In particular, $\{e=0\}$ is exponentially stable. Moreover, if $r_\chi>0$, then, for each $r\in(0,r_\chi)$, there is no $C>0$ for which \eqref{decay} holds for each initial condition $(e_0,e_1)\in\mathbb{R}^{2}$.
\item If $r_\chi\geq 1$, then, for each $\delta>0$, there exists an initial condition $(e_0,e_1)\in\mathbb{R}^{2}$ satisfying $\max_{j\in\{0,1\}}|e_j|\leq\delta$ such that the solution to \eqref{recurrence} satisfies $\limsup_{k\to+\infty}|e_k|>0$. In particular, $e_k$ does not converge to $0$ as $k\to+\infty$.
\end{itemize}
As a consequence, $\{e=0\}$ is exponentially stable if and only if all the poles of the closed-loop system belong to the open unit disc $\{z\in\mathbb{C} \, : \, |z|<1\}$.
\end{Proposition}

In the case where the origin is exponentially stable with $r_{\chi}>0$, writing $\lambda := -\ln (r)/h$, where $r\in (r_{\chi},1)$, we find that \eqref{decay} is equivalent to 
\begin{align}
|e_k| \leq C e^{-\lambda t_k} \max_{j\in \{0,1\}}|e_j| \qquad \forall k\in \mathbb{N}.
\end{align}
Since $r>r_\chi$, then $\lambda < \lambda_\chi$ where 
\begin{align}
\lambda_\chi := -\ln(r_\chi)/h.
\end{align}
We refer to $\lambda_\chi$ as the \textit{decay rate}. 

When $r_\chi <1$, we say that the characteristic polynomial $\chi(z)$ is \textit{Schur}.  
\begin{Proposition}
The characteristic polynomial $\chi(z)=z^2+c_1z+c_0$ is Schur if and only if $\chi(1)>0$, $\chi(-1)>0$ and $|c_0|<1$.
\end{Proposition}

Using Jury's test, we can derive necessary and sufficient conditions linking the controller gains (in the coefficients of $K(q)$), the system parameters and the sampling step to ensure exponential stability.

\subsection{PI controller}\label{sec_3}

The PI controller reads 
\begin{align}
u_k &:= - k_Pe_k - k_i I_k \quad \forall k\in \mathbb{N}, \label{PI_1} \\
I_k &:= I_{k-1} + he_k \quad \forall k\in \mathbb{N}^*, \qquad I_0 \in \mathbb{R},\label{PI_2} 
\end{align}
where $k_P\in \mathbb{R}$ is the proportional gain and $k_i\in \mathbb{R}$ is the integral gain. As a result, 
\begin{align}
u_{k+1}-u_k =&~ -k_Pe_{k+1}-k_i I_{k+1} + k_Pe_k + k_iI_k \nonumber \\
=&~ -(k_P+k_ih)e_{k+1}+k_Pe_k \quad \forall k\in \mathbb{N}. 
\end{align}
Hence, \eqref{controller} holds with 
\begin{align}
K(q) := (k_P+k_ih)q-k_P. \label{K_Pi}
\end{align}
The characteristic polynomial is therefore given by 
\begin{align}
\chi(q) &= (q-\bar{p})(q-1)+\bar{b}(k_P+k_ih)q-\bar{b}k_P \nonumber \\
&= q^2+(k_i\bar{b}h+k_P\bar{b}-1-\bar{p})q+\bar{p}-k_P\bar{b}. \label{C_pi}
\end{align}

Now that we have the characteristic polynomial, we can apply Jury's test to find the following result. 

\begin{proposition}\label{prop_stab_pi}
Consider the system \eqref{first}--\eqref{first_d2} under the PI controller \eqref{PI_1}-\eqref{PI_2}. Furthermore, let 
\begin{align}
c(x) := x \coth\left(\frac{x}{2}\right) \quad \forall x\in \mathbb{R}^*, \quad c(0) := 2. \label{c_def}
\end{align}
The origin $\{e=0\}$ is exponentially stable if and only if
\begin{align}
&bhk_i >0, \label{PI_cond_1} \\
&bh(2k_P+hk_i) < 2c(ph), \label{PI_cond_2} \\
&ph < bhk_P < c(ph). \label{PI_cond_3}
\end{align}
\hfill $\square$
\end{proposition}
\begin{proof}
By Jury's test, $\{e=0\}$ is exponentially stable if and only if
\begin{align}
&\bar{b} hk_i>0, \label{eq_PI1_1}\\
&\bar{b}(2k_P+hk_i)<2(1+\bar{p}), \label{eq_PI1_2}\\
&|\bar{p}-\bar{b} k_P|<1. \label{eq_PI1_3}
\end{align}
These are the well-known necessary and sufficient conditions for the stability of discrete-time first-order SISO LTI systems under the PI controller. The conditions \eqref{PI_cond_1}-\eqref{PI_cond_2}-\eqref{PI_cond_3} are these same conditions written in the original coefficients $p$ and $b$, instead of $\bar{p}$ and $\bar{b}$. 

Using the definition of $\bar{b}$ in \eqref{first_d2}, we conclude that \eqref{eq_PI1_1} is equivalent to \eqref{PI_cond_1}. 

Next, using \eqref{first_d2}, note that \eqref{eq_PI1_2} can be rewritten as 
\begin{align}
b\left[\int_0^h e^{p(h-\theta)}d\theta \right] (2k_P+hk_i) < 2(1+e^{ph}). \label{to_use_1}
\end{align}
If $p=0$, then \eqref{to_use_1} becomes
\begin{align}
bh(2k_P+hk_i)<4,
\end{align}
which is \eqref{PI_cond_2} since $c(0)=2$. If $p\neq 0$, then 
\begin{align}
\int_0^h e^{p(h-\theta)}d\theta =  \frac{e^{ph}-1}{p}. 
\end{align}
Hence, \eqref{to_use_1} becomes 
\begin{align}
b\frac{e^{ph}-1}{p}(2k_P+hk_i) < 2(1+e^{ph}).\label{to_use_2}
\end{align}
If $p>0$, then $e^{ph}-1>0$. If $p<0$, then $e^{ph}-1<0$. Hence, regardless of the sign of $p$, we have 
\begin{align}
\frac{e^{ph}-1}{p} >0. 
\end{align}
As a result, \eqref{to_use_2} becomes 
\begin{align}
b(2k_P+hk_i) &< 2p \frac{e^{ph}+1}{e^{ph}-1} \nonumber \\
&= 2p \coth \left(\frac{ph}{2}\right) \nonumber \\
&= \frac{2}{h}c(ph), 
\end{align}
which is \eqref{PI_cond_2}. 

Finally, \eqref{eq_PI1_3} can be rewritten as 
\begin{align}
-1+\bar{p} < \bar{b}k_P < 1+\bar{p}. \label{to_use_3}
\end{align}
If $p=0$, then \eqref{to_use_3} becomes 
\begin{align}
0 < bhk_P < 2,
\end{align}
which is \eqref{PI_cond_3}. If $p\neq 0$, then \eqref{to_use_3} becomes 
\begin{align}
-1+e^{ph} < b\frac{e^{ph}-1}{p}k_P < 1+e^{ph}.
\end{align}
Hence, 
\begin{align}
p<bk_P < \frac{c(ph)}{h}
\end{align}
which is \eqref{PI_cond_3}. 
\end{proof}

Additionally, since the characteristic polynomial is of degree $2$, we can derive an explicit relation between the poles of the closed-loop system and the gains $k_P$ and $k_i$. Indeed, writing \eqref{C_pi} as
$$
\chi(z)=(z-z_1)(z-z_2)=z^2-(z_1+z_2)z+z_1z_2,
$$
where $z_1$ and $z_2$ are the two poles of the closed-loop system, and identifying the coefficients with those of \eqref{C_pi}, directly yields the following result. 

\begin{proposition}\label{prop_roots}
Consider the system \eqref{first}--\eqref{first_d2} under the PI controller \eqref{PI_1}-\eqref{PI_2}. The poles of the closed-loop system are $z_1$ and $z_2$ (real, or complex conjugate) if and only if
\begin{align}
k_P&=\frac{\bar{p}-z_1z_2}{\bar{b}}, \label{eq_place_1}\\
k_i&=\frac{(1-z_1)(1-z_2)}{\bar{b} h}.
\label{eq_place_2}
\end{align}
\hfill $\square$
\end{proposition}

As a consequence, if we want to stabilize the origin at a prescribed decay rate $\lambda_\chi >0$, we can set $z_1 = e^{-\lambda_\chi h}$, then select any $z_2$ real with $|z_2|< z_1$, and finally chose the gains $k_P$ and $k_i$ such that \eqref{eq_place_1}-\eqref{eq_place_2} hold. This is the usual pole placement method. 

\subsection{iP controller}\label{ip_sec}

There are several ways to explain the iP controller in continuous time. One way is as follows. One starts by selecting some constant $\alpha \neq 0$, and then defining 
\begin{align}
F(t) := \dot{y}(t) - \alpha u(t).\label{F_def}
\end{align}
Rearranging the terms, we obtain the so-called \textit{ultra-local model} 
\begin{align}
\dot{y}(t) = F(t) + \alpha u(t). 
\end{align}
Note that this definition does not require us to know anything about the system, beyond the fact that it has a control input $u(t)$ and an output $y(t)$. The function $F$ lumps together everything that is unknown about the system. For example, in the particular case of the first-order system \eqref{first}, we have 
\begin{align}
F(t) = py(t) + (b-\alpha)u(t). 
\end{align}
The iP controller then takes, in continuous time, the form 
\begin{align}
u(t) := -\frac{1}{\alpha}\left(\hat{F}(t)+K_P e(t)\right),\label{ip_law}
\end{align}
where $\hat{F}$ is an estimate of $F$, and $K_P \neq 0$ is a controller gain. The closed-loop system then becomes 
\begin{align}
\dot{y}(t) = -K_Pe(t) + F(t)-\hat{F}(t),
\end{align}
and, using the fact that $y^*$ is constant, we get 
\begin{align}
\dot{e}(t) = -K_P e(t) + F(t)-\hat{F}(t).
\end{align}
If the estimate $\hat{F}$ were perfect, in the sense that $F-\hat{F} \equiv 0$, then we would have $\dot{e}(t) = -K_P e(t)$, which would yield exponential stability of $\{e=0\}$ in continuous time if and only if $K_P >0$.

The discrete form of \eqref{ip_law} is
\begin{align}
u_k := -\frac{1}{\alpha} \left(\hat{F}_k + K_P e_k\right) \quad \forall k\in \mathbb{N}.\label{ip_1} 
\end{align}
The simplest way to design $\hat{F}_k$ is
\begin{align}
\hat{F}_k := \frac{y_k-y_{k-1}}{h}-\alpha u_{k-1} \quad \forall k\in \mathbb{N}^*, \qquad \hat{F}_0 \in \mathbb{R}.\label{ip_2}
\end{align}
The motivation behind this design of $\hat{F}_k$ is that, formally speaking, if $h$ is "very small", then by \eqref{F_def} we have $\hat{F}_k \approx F_{k-1}$ for $k\in \mathbb{N}^*$, and if $F$ "does not vary too quickly", then $F_k-\hat{F}_k\approx F_k-F_{k-1} \approx 0$ for $k\in \mathbb{N}^*$.

\section{Stability analysis under the iP controller}\label{sec_inp}

\subsection{Main results}

The iP controller \eqref{ip_1}-\eqref{ip_2} can be rewritten as  
\begin{align*}
u_{k+1} =&~ -\frac{1}{\alpha}\left(\hat{F}_{k+1}+K_Pe_{k+1}\right) \nonumber \\
=&~ -\frac{1}{\alpha}\left( \frac{y_{k+1}-y_k}{h}-\alpha u_k + K_P e_{k+1}\right) \nonumber \\
=&~ -\frac{1}{\alpha}\left( \frac{e_{k+1}-e_k}{h}-\alpha u_k + K_P e_{k+1}\right) \quad \forall k\in \mathbb{N},
\end{align*}
where we have used the fact that $y_{k+1}-y_k = e_{k+1}-e_k$ since $y^*$ is constant. Hence, 
\begin{align}
u_{k+1}-u_k = -\frac{K_P}{\alpha}e_{k+1} - \frac{e_{k+1}-e_{k}}{\alpha h} \quad \forall k\in \mathbb{N}. 
\end{align}
As a consequence, \eqref{controller} holds with 
\begin{align}
K(q) := \left(\frac{K_P}{\alpha}+\frac{1}{\alpha h}\right)q - \frac{1}{\alpha h}. \label{K_ip}
\end{align}
The corresponding characteristic polynomial is given by 
\begin{align}
\chi(q) =&~ (q-\bar{p})(q-1)+\bar{b} \left[\left(\frac{K_P}{\alpha}+\frac{1}{\alpha h}\right)q-\frac{1}{\alpha h} \right] \nonumber \\
=&~ q^2+\left[\bar{b}\left(\frac{K_P}{\alpha}+\frac{1}{\alpha h}\right)-1-\bar{p}\right]q + \bar{p}-\frac{\bar{b}}{\alpha h}. \label{C_ip}
\end{align}
Comparing \eqref{C_pi} with \eqref{C_ip}, we find that the PI and iP controllers yield the same characteristic polynomial for the closed-loop system if and only if 
\begin{align}
\bar{p}-k_P\bar{b} &= \bar{p}-\frac{\bar{b}}{\alpha h}, \nonumber \\
k_i\bar{b}h+k_P\bar{b}-1-\bar{p} &= \bar{b}\left(\frac{K_P}{\alpha}+\frac{1}{\alpha h}\right)-1-\bar{p}.
\end{align}
That is, if and only if 
\begin{align}
k_P = \frac{1}{\alpha h}, \qquad k_i = \frac{K_P}{\alpha h}. \label{equivalence}
\end{align}
The equivalence between the discrete-time PI and iP controllers through \eqref{equivalence} has already been observed in \cite{DN10}. However, in that reference, this equivalence was used to explain the ubiquity of the PI controller in practice, given that it is equivalent to an iP controller whose performance may be intuitively explained by the heuristics presented in Section \ref{ip_sec}. To the best of our knowledge, this equivalence has never been used in the opposite direction, namely, to study the stability of the iP controller using existing results for the PI controller. In particular, we are interested in the following question:
\begin{center}
\fbox{\parbox{0.4\textwidth}{
\centering
\textit{If the iP and PI are two parametrizations of the same controller, why use an iP rather than a PI?}
}}
\end{center}
This paper is a first step towards answering this question.

Using Proposition \ref{prop_stab_pi}, we can establish the 
\begin{Theorem}\label{thm_1}
Consider the system \eqref{first}--\eqref{first_d2} under the iP controller \eqref{ip_1}-\eqref{ip_2}. Furthermore, let 
\begin{align}
\rho := \frac{b}{\alpha}, \label{rho_def}
\end{align}
and $c$ be defined according to \eqref{c_def}. The origin $\{e=0\}$ is exponentially stable if and only if 
\begin{align}
&\rho K_P>0, \label{43}\\
&\rho (2+hK_P)<2 c(ph), \label{44} \\
&ph<\rho<c(ph). \label{45}
\end{align}
\end{Theorem}
\begin{proof}
Using Proposition \ref{prop_stab_pi} and the equivalence of the PI and iP controllers through \eqref{equivalence}, we conclude that the origin $\{e=0\}$ under the iP controller is exponentially stable if and only if 
\begin{align}
&bh\frac{K_P}{\alpha h} >0, \label{iP_to_use1} \\
&bh\left(\frac{2}{\alpha h}+h\frac{K_P}{\alpha h}\right) < 2 c(ph), \label{iP_to_use2} \\
&ph < bh\frac{1}{\alpha h} < c(ph). \label{iP_to_use3}
\end{align}
Clearly, \eqref{iP_to_use1} is \eqref{43}, \eqref{iP_to_use2} is \eqref{44}, and \eqref{iP_to_use3} is \eqref{45}. 
\end{proof}

Our second main result is the
\begin{theorem}\label{thm:small_h}
Consider the system \eqref{first}--\eqref{first_d2} under the iP controller \eqref{ip_1}-\eqref{ip_2}, and let $\rho$ be defined according to \eqref{rho_def}. Furthermore, assume that
\begin{align}
0<\rho<2,
\quad
K_P>0.
\end{align}
Then, there exists $h_0>0$ such that, for every $h\in(0,h_0)$, the characteristic polynomial $\chi$ in \eqref{C_ip} has two real roots $z_f(h)$ and $z_s(h)$ satisfying
\begin{align}
z_f(h)=1-\rho+O(h) \qquad \text{(\textit{fast} pole)},
\end{align}
and
\begin{align}
z_s(h)=1-hK_P+O(h^2) \qquad \text{(\textit{slow} pole)}.
\end{align}
Moreover, after possibly decreasing $h_0$, one has
\begin{align}
|z_f(h)|<z_s(h)<1,
\qquad
\forall h\in(0,h_0).
\end{align}
Consequently, the spectral radius is
\begin{align}
r_\chi(h)=z_s(h),
\end{align}
and the corresponding decay rate is
\begin{align}
\lambda_\chi(h)=K_P+O(h).
\end{align}
In particular,
\begin{align}
\lim_{h\to0^+}\lambda_\chi(h)=K_P.
\end{align}
\hfill $\square$
\end{theorem}
\begin{proof}
We first write the characteristic polynomial in \eqref{C_ip} in a form that is convenient for studying the limit as $h\to0$.

Introduce
\begin{align}
\varphi(h):=\frac{e^{ph}-1}{ph} \ \ \text{if $ph\neq 0$}, \qquad \varphi := 1 \ \ \text{if $ph=0$}.
\end{align}
Then, from \eqref{first_d2}, we have 
\begin{align}
\bar b=bh\varphi(h),
\end{align}
and therefore
\begin{align}
\frac{\bar b}{\alpha h}=\rho\varphi(h),
\qquad
\frac{\bar bK_P}{\alpha}
=\rho hK_P\varphi(h).
\end{align}
Using \eqref{C_ip}, we consequently have
\begin{align*}
\chi(z)
=
z^2+
\left[
\rho\varphi(h)(1+hK_P)-1-e^{ph}
\right]z
+e^{ph}-\rho\varphi(h).
\end{align*}
We write $\chi_h(z):=\chi(z)$ to insist on the dependence of $\chi$ on $h$. Since $e^{ph}$ and $\varphi(h)$ are analytic in $h$ around $h=0$, the function
\begin{align}
\Phi(z,h)=\chi_h(z)
\end{align}
is analytic in $(z,h)$ around $h=0$. Moreover,
\begin{align}
e^{ph}=1+ph+O(h^2),
\end{align}
and
\begin{align}
\varphi(h)=1+\frac{p}{2}h+O(h^2).
\end{align}
Hence
\begin{align}
\chi_h(z)
&=
z^2+(\rho-2)z+1-\rho+O(h).
\end{align}
Therefore, at $h=0$,
\begin{align}
\chi_0(z)
&=
z^2+(\rho-2)z+1-\rho\\
&=
(z-1)(z-(1-\rho)).
\end{align}
Thus, the limiting characteristic polynomial has two distinct roots,
\begin{align}
z=1-\rho
\end{align}
and
\begin{align}
z=1.
\end{align}
We first study the root $1-\rho$. Since
\begin{align}
\Phi(1-\rho,0)=0,
\end{align}
we compute
\begin{align}
\partial_z \Phi(1-\rho,0)
&=
2(1-\rho)+\rho-2\\
&=-\rho.
\end{align}
The assumption $\rho>0$ therefore gives
\begin{align}
\partial_z\Phi(1-\rho,0)\neq0.
\end{align}
The implicit function theorem implies that there exist $h_f>0$ and a unique analytic function $z_f(h)$ defined for $h\in(-h_f,h_f)$ such that
\begin{align}
z_f(0)=1-\rho
\end{align}
and
\begin{align}
\Phi(z_f(h),h)=0 \quad \forall h\in (-h_f,h_f).
\end{align}
Since $z_f$ is analytic in a neighborhood of $h=0$ (because $\Phi$ is analytic around $h=0$), it admits a Taylor expansion of the form
\begin{align}
z_f(h)
&=z_f(0)+z_f'(0)h+O(h^2)\\
&=1-\rho+O(h).
\end{align}
Since $0<\rho<2$, we have
\begin{align}
|1-\rho|<1.
\end{align}
Hence, by continuity of $h\mapsto z_f(h)$, there exist $\varepsilon>0$ and $h_1>0$ such that
\begin{align}
|z_f(h)|<1-\varepsilon
\qquad
\forall h\in(0,h_1).
\end{align}
Thus, the fast pole remains uniformly separated from the unit circle for sufficiently small $h$.

We now study the root $z=1$. The previous argument does not directly give its first-order behavior because this root approaches the unit circle as $h$ tends to zero. We therefore introduce the rescaling
\begin{align}
G(s,h)=\frac{1}{h}\chi_h(1+hs) \quad \forall (h,s)\in \mathbb{R}^*\times \mathbb{R}.
\end{align}
Substituting the expression of $\chi_h$ gives
\begin{align}
hG(s,h) =&~ (1+hs)^2 \nonumber \\
&~+\left[\rho \varphi(h)(1+hK_P)-1-e^{ph}\right](1+hs) \nonumber \\
&~+e^{ph}-\rho \varphi(h) \nonumber \\
=&~ 1+h^2s^2+2hs \nonumber \\
&~+\rho\varphi(h)(1+hK_P)(1+hs) \nonumber \\
&~-(1+e^{ph})(1+hs)+e^{ph}-\rho \varphi(h) \nonumber \\
=&~ \rho \varphi(h) h (s+K_P+hsK_P)+hs(1+hs-e^{ph}). \nonumber 
\end{align}
Hence, 
\begin{align*}
G(s,h) = \rho \varphi(h) (s+K_P+hsK_P)+s(1+hs-e^{ph}).
\end{align*}
Since $\varphi$ and the exponential function are analytic around $h=0$, the right-hand side defines an analytic function of $(s,h)$ in a neighborhood of $h=0$. We henceforth use $G$ to denote this analytic extension, which is obtained by evaluating the above expression also at $h=0$. In particular, since $\varphi(0)=1$, we have
\begin{align}
G(s,0) &=\rho(s+K_P).
\end{align}
Consequently,
\begin{align}
G(-K_P,0)=0.
\end{align}
Moreover,
\begin{align}
\partial_sG(-K_P,0)=\rho\neq0.
\end{align}
The implicit function theorem therefore gives $h_s>0$ and a unique analytic function $s(h)$ defined for $h\in(-h_s,h_s)$ such that
\begin{align}
s(0)=-K_P
\end{align}
and
\begin{align}
G(s(h),h)=0 \quad \forall h\in (-h_s,h_s).
\end{align}
Hence,
\begin{align}
s(h)=-K_P+O(h).
\end{align}
Returning to the original variable $z$, we have
\begin{align}
z_s(h)
&=
1+hs(h)\\
&=
1-hK_P+O(h^2).
\end{align}
Since $K_P>0$, it follows that
\begin{align}
z_s(h)<1
\end{align}
for all sufficiently small positive $h$. Moreover, since $z_s(h)$ tends to $1$, we also have
\begin{align}
z_s(h)>0
\end{align}
for all sufficiently small positive $h$. Therefore,
\begin{align}
|z_s(h)|=z_s(h).
\end{align}

We now compare the two poles. From the first part of the proof,
\begin{align}
|z_f(h)|<1-\varepsilon
\end{align}
for sufficiently small $h$, whereas
\begin{align}
z_s(h)\longrightarrow1.
\end{align}
Consequently, after possibly decreasing $h_0$, we have
\begin{align}
|z_f(h)|<z_s(h)<1
\qquad
\forall h\in(0,h_0).
\end{align}
Thus, for every $h\in(0,h_0)$, the spectral radius is determined by the slow pole
\begin{align}
r_\chi(h)
&=
\max\{|z_f(h)|,|z_s(h)|\}\\
&=
z_s(h).
\end{align}
Using the expansion of the slow pole, we obtain
\begin{align}
r_\chi(h)
&=
1-hK_P+O(h^2).
\end{align}
It remains to study the decay rate. By definition,
\begin{align}
\lambda_\chi(h)
&=
-\frac{1}{h}\ln \left(r_\chi(h)\right).
\end{align}
Since $r_\chi(h)=1-hK_P+O(h^2)$, the expansion
\begin{align}
\ln(1+x)=x+O(x^2)
\end{align}
with
\begin{align}
x=-hK_P+O(h^2)
\end{align}
gives
\begin{align}
\ln \left( r_\chi(h)\right)
&=
-hK_P+O(h^2).
\end{align}
Consequently,
\begin{align}
\lambda_\chi(h)
&=
-\frac{1}{h}
\left(
-hK_P+O(h^2)
\right)\\
&=
K_P+O(h).
\end{align}
Therefore,
\begin{align}
\lim_{h\to0^+}\lambda_\chi(h)=K_P,
\end{align}
which concludes the proof.
\end{proof}

\subsection{Interpretation of the main results}

The iP controller, and more generally, iPID controllers, are usually presented in the literature as being model-free, in the sense that their tuning does not require knowledge of the system dynamics. Certainly, the iP law~\eqref{ip_1}-\eqref{ip_2} uses only the measured output, the previous input, and the three numbers $\alpha$, $K_P$ and $h$. Neither $p$ nor $b$ enters its implementation. However, the same is true for the PI controller: we can choose any values of $k_P$ and $k_i$. Yet, to ensure stability, the tuning requires some knowledge of the system parameters $p$ and $b$, either for the PI controller (through Proposition \ref{prop_stab_pi}) or for the iP controller (through Theorem \ref{thm_1}). In particular, the condition \eqref{43} requires the designer to know the control direction $\sign(b)$. Indeed, if $\sign(\alpha) = \sign(b)$, then $\rho >0$, and we must have $K_P>0$. Otherwise, if $\sign(\alpha)=-\sign(b)$, then $\rho < 0$ and we must instead have $K_P<0$. Hence, we distinguish between two cases:

\begin{itemize}
    \item Case 1 ($\rho <0$ and $K_P < 0$). The left inequality in \eqref{45} can hold only if
\begin{equation}\label{int:wrong_sign}
ph<\rho<0,
\end{equation}
that is, only if $p<0$ and the sampling period satisfies
\begin{equation}\label{int:wrong_sign_h}
h>\frac{\rho}{p}.
\end{equation}
Hence, we need knowledge of $\rho/p$ to ensure that the sampling period is not chosen too small. The right inequality in \eqref{45} then holds for free, since $c(x)\geq 2>0$ for all $x\in\mathbb{R}$. Condition \eqref{44} also holds as long as $h|K_P|\leq 2$, otherwise, it requires $|\rho| (h|K_P|-2)<2c(ph)$. This case reveals some limitations of the heuristic in Section \ref{ip_sec}, which suggests always choosing $K_P>0$ regardless of $\alpha$, since in the limit $h\to 0$, $\alpha$ disappears and one obtains $\dot{e}=-K_Pe$. This explains the restriction \eqref{int:wrong_sign_h} on the sampling period in this case. 
\item Case 2 ($\rho >0$ and $K_P >0$). Since $c(x) \geq 2$ for all $x\in \mathbb{R}$, the condition \eqref{44} can be enforced without knowledge of $p$ by letting 
$$\rho (2+hK_P) < 4.$$
However, this requires knowledge of an upper bound on $b$, since $\rho = b/\alpha$. For the third condition \eqref{45}, it then holds for free if $p \leq 0$, so no knowledge of the size of $p$ is needed (note, however, that in this case, the traditional PI controller does not require knowledge of $p$ either). However, if $p>0$, then we must choose the sampling period sufficiently small, so that
\begin{align}
h < \frac{\rho}{p}. 
\end{align}
Interestingly, for the PI controller, instead of choosing a small sampling period, we would need to choose a sufficiently large gain $k_P$ to satisfy \eqref{PI_cond_3}. Hence, by reparametrizing the PI controller as an iP, we have transferred the problem from selecting a high gain to using a fast sampling rate to ensure stability. 
\end{itemize}

Obviously, since we are dealing with linear systems, we do not distinguish between local and global exponential stability: if $\{e=0\}$ is not exponentially stable, then it is not locally exponentially stable. 

In summary, the practical difference between the PI and iP controllers lies in the tuning variable. With a PI controller, the practitioner adjusts the gains at a fixed sampling period. With the iP controller, the practitioner may keep $\alpha$ and $K_P$ fixed and decrease the sampling period.

It is interesting to note that the two tuning philosophies also differ in terms of convergence rate, in light of Theorem \ref{thm:small_h}. For the PI controller, if we keep $k_P$ and $k_i$ fixed and decrease the sampling period, the decay rate converges to that of the continuous-time closed loop, determined by the roots of $s^2+(bk_P-p)s+bk_i$, which depend on $p$ and $b$. In contrast, for the iP controller with $\alpha$ and $K_P$ fixed and $0<\rho<2$, the decay rate converges to $K_P$ as $h\to0$. Hence, in this limiting sense, the dominant decay rate can be prescribed by choosing $K_P$, using only the sign of $b$ and an upper bound on $|b|$, without knowledge of $p$. Certainly, this analysis does not take into consideration measurement noise, which would enforce some lower-bound on $h$. 

\section{Numerical experiments}\label{sec:numerics}

All experiments use the plant \eqref{first} with $y(0)=0$ and the reference $y^*=1$. The control input is held constant on each sampling interval, and the output between sampling instants is obtained from the exact solution of \eqref{first}. Both controllers are implemented exactly as in \eqref{PI_1}-\eqref{PI_2} and \eqref{ip_1}-\eqref{ip_2}, with the initializations $I_0=0$ and $\hat{F}_0=0$. The regulation error $|y(t)-y^*|$ is plotted in logarithmic scale. On such a plot, an exponentially decaying error eventually becomes a straight line whose slope is minus the decay rate, so that the convergence rate can be read directly from the trajectories. The decay rates quoted below and in the tables are nevertheless not estimated from the curves: they are computed from the roots of the characteristic polynomial \eqref{C_pi} or \eqref{C_ip} as $\lambda_\chi=-\ln(r_\chi)/h$.

\subsection{Fixed parameters and decreasing sampling period}

The first experiment uses the open-loop unstable plant corresponding to $p=b=1$. The PI gains are $k_P=7$ and $k_i=5$, so that the continuous-time closed-loop characteristic polynomial is
\begin{equation}\label{num:ct1}
s^2+(bk_P-p)s+bk_i=s^2+6s+5=(s+1)(s+5).
\end{equation}
Its slowest root is $-1$, and the continuous-time decay rate is $\lambda_{\mathrm{PI,ct}}=1$. The iP controller uses $\alpha=1$ and $K_P=2$, so that $\rho=1$ and the assumptions of Theorem~\ref{thm:small_h} hold. All parameters are kept fixed while the sampling period decreases.

\begin{figure*}[t]
\centering
\includegraphics[width=0.9\textwidth]{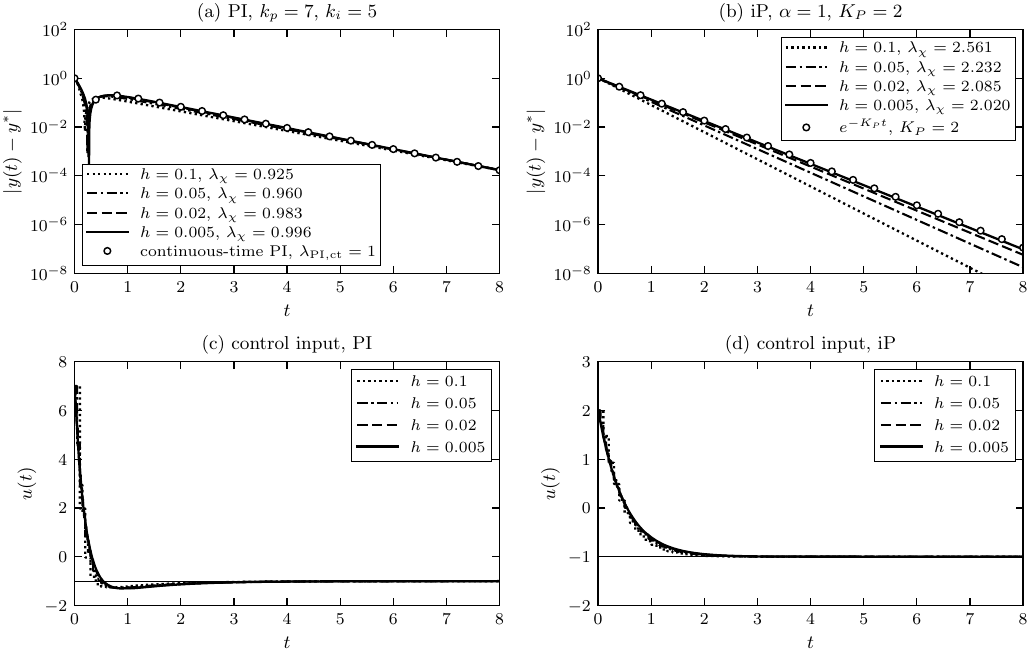}
\caption{Experiment~1: $p=b=1$, fixed parameters, decreasing $h$. (a) Regulation error of the PI loop ($k_P=7$, $k_i=5$); circles: continuous-time PI loop, whose slowest root is $-1$. (b) Regulation error of the iP loop ($\alpha=1$, $K_P=2$); circles: $e^{-K_Pt}$. (c), (d) Control inputs; thin line: $-py^*/b=-1$. The legends give the exact decay rates $\lambda_\chi$ computed from the closed-loop characteristic polynomials.}
\label{fig:exp1}
\end{figure*}

Fig. \ref{fig:exp1}(a) shows that the regulation error under the PI controller converge, as $h$ decreases, to the regulation error of the continuous-time PI loop (circles). After a short transient, all curves become straight lines with slope $-1$, the slowest root of \eqref{num:ct1}. The notch near $t\approx0.27$ is a sign change of the error, caused by the overshoot: the continuous-time error is $0.5e^{-t}-1.5e^{-5t}$, so the fast mode dominates initially and the slow mode afterwards. Table \ref{tab:exp1} quantifies this convergence. The decay rate of the PI loop increases from $0.925$ at $h=0.1$ to $0.999$ at $h=0.001$ and tends to $\lambda_{\mathrm{PI,ct}}=1$. This limit is determined by the gains and by the plant parameters $p$ and $b$ through \eqref{num:ct1}.

Fig. \ref{fig:exp1}(b) shows the regulation error under the iP controller. We can observe straight lines almost from $t=0$, because, with the initialization $\hat F_0=0$ and $y(0)=0$, the fast mode is only weakly excited. Their slopes decrease towards $K_P=2$ as $h$ decreases: the exact decay rate is $2.561$ at $h=0.1$, $2.020$ at $h=0.005$, and $2.004$ at $h=0.001$. This is the content of Theorem \ref{thm:small_h}: with $\alpha$ and $K_P$ fixed, the limiting decay rate is $K_P$, a parameter that is chosen directly without knowledge of $p$ and $b$. 

The control inputs in Fig. \ref{fig:exp1}(c)--(d) converge to $-py^*/b=-1$, the input that holds the plant at $y=1$. The PI input starts at $u_0=k_P|e_0|=7$, which reflects the proportional action on the initial error, whereas the iP input starts at $u_0=K_P|e_0|/\alpha=2$. As $h$ decreases, the iP input approaches $-1+3e^{-2t}$, which is exactly the input for which the plant follows $\dot e=-K_Pe$ from $y(0)=0$. In both cases the inputs are bounded and depend only weakly on $h$ in this noise-free setting.

\subsection{Small sampling period and different plants}

The second experiment keeps $h=0.01$ and all controller parameters fixed, and varies the open-loop pole: $p\in\{-2,0,1,2\}$ with $b=1$. All closed loops are exponentially stable. For $p=2$, the iP controller requires $h<\rho/p=0.5$, which holds.

\begin{figure*}[t]
\centering
\includegraphics[width=0.9\textwidth]{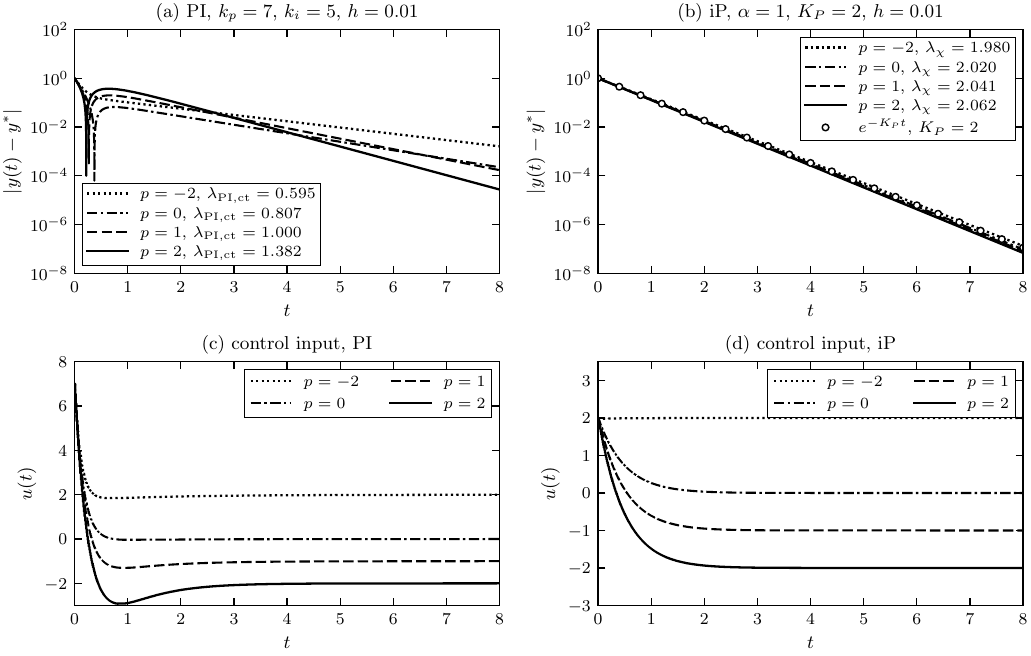}
\caption{Experiment~2: $h=0.01$, $b=1$, $p\in\{-2,0,1,2\}$, with the same controller parameters for every plant. (a) PI ($k_P=7$, $k_i=5$); the legend gives the slowest root of $s^2+(bk_P-p)s+bk_i$. (b) iP ($\alpha=1$, $K_P=2$); the legend gives the exact decay rates; circles: $e^{-K_Pt}$. (c), (d) Control inputs.}
\label{fig:exp2}
\end{figure*}

In Fig. \ref{fig:exp2}(a), the PI errors decay with clearly different slopes. Their decay rates follow the slowest root of $s^2+(7-p)s+5$, which equals $0.595$, $0.807$, $1$ and $1.382$ for the four plants. The exact sampled-data rates are $0.593$, $0.803$, $0.991$ and $1.354$ (Table \ref{tab:exp2}). With fixed gains, the convergence rate of the PI loop is therefore a property of the plant as much as of the controller. Prescribing it requires knowledge of $p$ and $b$, for instance through the pole-placement formulas \eqref{eq_place_1}--\eqref{eq_place_2}.

In Fig. \ref{fig:exp2}(b), the four iP errors lie on the same straight line $e^{-K_Pt}$. Their exact decay rates are $1.980$, $2.020$, $2.041$ and $2.062$. They differ from $K_P=2$ by an amount of order $h$ that grows with $p$, and they converge to $K_P$ for every plant as $h$ decreases further. This is the most important observation of this section. With the same $\alpha$ and $K_P$, the decay rate of the iP loop is essentially the same for an open-loop stable, integrating or unstable plant, without any information on $p$ having been used.

The control inputs are depicted in Fig. \ref{fig:exp2}(c)-(d). For both controllers, the input converges to the plant-dependent value $-py^*/b$, which equals $2$, $0$, $-1$ and $-2$. The iP input, however, adapts to the plant so that the error follows the same exponential in every case. For $p=-2$, the open-loop response of the plant to the constant input $u=2$ is already $y=1-e^{-2t}$, so the iP input remains almost constant, within $0.02$ of its initial value. The PI input, in contrast, starts with the same proportional kick $u_0=7$ for every plant and then follows plant-dependent transients.

\subsection{Wrong control direction: $\rho<0$}

\begin{figure*}[t]
\centering
\includegraphics[width=0.9\textwidth]{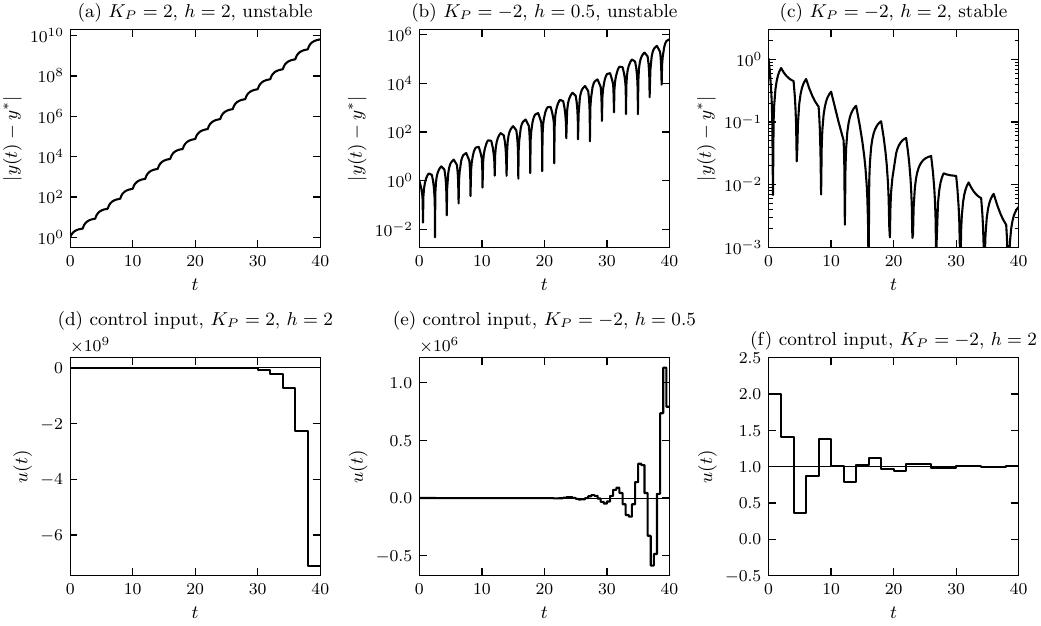}
\caption{Experiment~3: iP controller with $\alpha=-1$ ($\rho=-1$) for the open-loop stable plant $p=-1$, $b=1$. Top: regulation error; bottom: control input (thin line: $-py^*/b=1$). (a), (d) $K_P=2$, $h=2$: unstable for every $h$, since $\rho K_P<0$. (b), (e) $K_P=-2$, $h=0.5$: unstable, since $h<\rho/p=1$. (c), (f) $K_P=-2$, $h=2$: stable, with decay rate $\lambda_\chi=0.142$.}
\label{fig:exp3}
\end{figure*}

The third experiment uses the open-loop stable plant corresponding to $p=-1$, $b=1$, and an iP controller with $\alpha=-1$, so that $\rho=b/\alpha=-1$. By Theorem~\ref{thm_1}, condition~\eqref{43} then requires $K_P<0$, and the left inequality in~\eqref{45}, $ph<\rho$, requires
\begin{equation}
h>\frac{\rho}{p}=1.
\end{equation}
For this example, \eqref{44} and the right inequality in~\eqref{45} hold for all $h$. The iP loop is therefore unstable for every sampling period when $K_P>0$, and stable if and only if $h>1$ when $K_P=-2$. A sweep over $h\in[0.01,10]$ confirmed both statements. Fig. \ref{fig:exp3} shows three representative cases.

With $K_P=2$ and $h=2$ (Fig. \ref{fig:exp3}(a),(d)), the closed loop has the real pole $z=3.115$. The error and the input diverge. The equivalence \eqref{equivalence} explains why no sampling period can help: the equivalent PI gains satisfy $bk_i=\rho K_P/h<0$, so the integral action has the wrong sign. Then $\chi(1)=\bar bk_ih<0$, and $\chi$ has a real root larger than one for every $h$.

With $K_P=-2$, the integral action has the correct sign, $bk_i=\rho K_P/h>0$, but the proportional action does not, since $bk_P=\rho/h=-1/h<0$. The loop therefore contains positive proportional feedback, and stability requires that the plant's own damping dominate it: $|bk_P|<|p|$, that is, $h>1$. For $h=0.5$ (Fig. \ref{fig:exp3}(b),(e)), $bk_P=-2$, the poles form a complex pair of modulus $1.18$, and the error and the input diverge in an oscillatory manner. For $h=2$ (Fig. \ref{fig:exp3}(c),(f)), $bk_P=-0.5$, the poles have modulus $0.753$, and the loop is stable: the error decays and the input converges to $-py^*/b=1$. The decay rate is only $\lambda_\chi=0.142$, however, much slower than the open-loop rate $|p|=1$ of the plant, and the convergence is oscillatory.

This experiment shows that the small-sampling-period reasoning breaks down completely when the sign of $\alpha$ is wrong. As $h$ decreases, one pole tends to $1-\rho=2$ whatever the sign of $K_P$. For instance, it equals $2.024$ for $K_P=2$ and $1.944$ for $K_P=-2$ at $h=0.01$. For $K_P<0$, the other pole also leaves the unit disc, since it behaves like $1-hK_P>1$. The heuristic according to which the error obeys $\dot e\approx-K_Pe$ for small $h$ therefore fails. The only stable configurations with $\rho<0$ require an open-loop stable plant, a negative $K_P$, and a sampling period larger than $\rho/p$, which presupposes knowledge of $p$. Even then, the resulting loop is slow. The control direction must be known for the iP controller, as for the PI controller.

% ---------------------------------------------------------------------
\begin{table}
\centering
\caption{Experiment~1 ($p=b=1$): exact decay rates $\lambda_\chi(h)=-\ln(r_\chi(h))/h$ of the PI loop ($k_P=7$, $k_i=5$) and of the iP loop ($\alpha=1$, $K_P=2$).}
\label{tab:exp1}
\begin{tabular}{c|cc}
\hline
$h$ & $\lambda_{\mathrm{PI}}(h)$ & $\lambda_{\mathrm{iP}}(h)$ \\
\hline
0.2 & 0.8671 & 5.4204 \\
0.1 & 0.9249 & 2.5612 \\
0.05 & 0.9596 & 2.2319 \\
0.02 & 0.9831 & 2.0846 \\
0.01 & 0.9914 & 2.0411 \\
0.005 & 0.9957 & 2.0203 \\
0.001 & 0.9991 & 2.0040 \\
\hline
$h\to0^+$ & $\lambda_{\mathrm{PI,ct}}=1$ & $K_P=2$ \\
\hline
\end{tabular}
\end{table}

\begin{table}
\centering
\caption{Experiment 2 ($b=1$, $h=0.01$): slowest root of $s^2+(bk_P-p)s+bk_i$ and exact decay rates, with the same parameters for every plant (PI: $k_P=7$, $k_i=5$; iP: $\alpha=1$, $K_P=2$).}
\label{tab:exp2}
\setlength{\tabcolsep}{4pt}
\begin{tabular}{c|cc|c}
\hline
$p$ & $\lambda_{\mathrm{PI,ct}}$ & $\lambda_{\mathrm{PI}}(h=0.01)$ & $\lambda_{\mathrm{iP}}(h=0.01)$ \\
\hline
$-2$ & 0.5949 & 0.5933 & 1.9799 \\
$0$ & 0.8074 & 0.8032 & 2.0203 \\
$1$ & 1.0000 & 0.9914 & 2.0411 \\
$2$ & 1.3820 & 1.3536 & 2.0624 \\
\hline
\end{tabular}
\end{table}

\section{Conclusion and research perspectives}
This paper has compared discrete-time PI and iP controllers for a first-order linear plant under sampled-data implementation. At every fixed sampling period, the two controllers generate the same closed loops, and the exact stability conditions of the PI loop yield necessary and sufficient conditions for the iP loop. The two controllers differ in the tuning they suggest. With fixed gains, the decay rate of the PI loop tends, as the sampling period decreases, to a value that depends on the plant. With fixed $\alpha$ and $K_P$, the decay rate of the iP loop tends to $K_P$, provided that $\alpha$ has the sign of the input gain and does not underestimate its magnitude by a factor of two or more.

The author is currently investigating general discrete-time iPID controllers applied to second-order control systems.

\bibliographystyle{IEEEtran}  
\bibliography{references}     

\end{document}